\documentclass{elsart3p}

\usepackage{graphicx,epsfig}

\begin{document}
\begin{frontmatter}
\title{Reconstruction of a  Radiation Point Source's Radial Location Using Goodness-of-Fit Test on Spectra Obtained from an HPGe Detector}

\author[TUNL,UNC]{L.T. Evans},
\ead{ltevans@unc.edu}
\author[TUNL,UNC]{K. Andre},
\ead{kali@unc.edu}
\author[TUNL]{R. De},
\ead{risingstar49@aol.com}
\author[TUNL,UNC]{R. Henning\corauthref{cor}},
\ead{rhenning@physics.unc.edu}
\corauth[cor]{Corresponding Author}
\author[TUNL,UNC]{E.D. Morgan}
\ead{edmorgan@email.unc.edu}

\address[TUNL]{Triangle University Nuclear Laboratories,
Durham, NC, USA}
\address[UNC]{University of North Carolina, Chapel Hill, NC, USA}
\begin{abstract}
High purity germanium (HPGe) detectors are ubiquitous in nuclear physics experiments and are also used in numerous low radioactive background
detectors. The effect of the position of $^{60}$Co and $^{137}$Cs point
sources on the shape of spectra were studied with Monte Carlo
and HPGe detector measurements. We briefly confirm previous work on the position dependence of relative heights of peaks. Spectra taken with the radiation
sources placed at locations around the detector were then compared using the
Kolmogorov-Smirnov (K-S) goodness-of-fit test. We demonstrate that with this method the Compton continuum spectral shape has good sensitivity to the radial location of a point-source, but poor angular resolution. We conclude with a study of the position reconstruction accuracy as a function of the number of counts from the source.
\end{abstract}
\end{frontmatter}

\section{Introduction}

High Purity Germanium (HPGe) detectors are used for $\gamma$-ray detection and spectroscopy in nuclear physics experiments, including low radioactive background detector systems, such as the proposed {\sc Majorana} experiment~\cite{majo}. Others have shown and explained how a detected spectrum shape can vary noticeably based on the position of a $\gamma$-emitting radiation source relative to the detector~\cite{bak, helmer, knoll, lich}. We show that the Kolmogorov-Smirnov (K-S) test is sensitive to these changes as well. We also show that the radial position of nearby $\gamma$-emitting radiation sources can be accurately reconstructed using the K-S test to compare spectra taken at different locations.

\section{Experiment}

For our measurements, we used an EG\&G Ortec HPGe detector model P40621A with a germanium crystal of length 93 mm and a diameter of 90 mm. It has an active volume of 582 cm$^3.$ The detector's cryostat has a 1.0 mm aluminum endcap.

\begin{figure}
\centering
\includegraphics[width=0.3\textwidth]{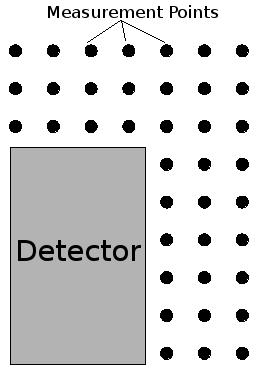}
\caption{Spectra were taken by moving the $^{60}$Co or $^{137}$Cs (in Monte Carlo simulation) source along a grid of resolution 2~mm (not to scale). The grid was marked on a flat acrylic sheet that was mounted to the detector cryostat (the gray box).}\label{fig:detectordiagram}
\end{figure}

\begin{figure}
\centering
\includegraphics[width=0.49\textwidth]{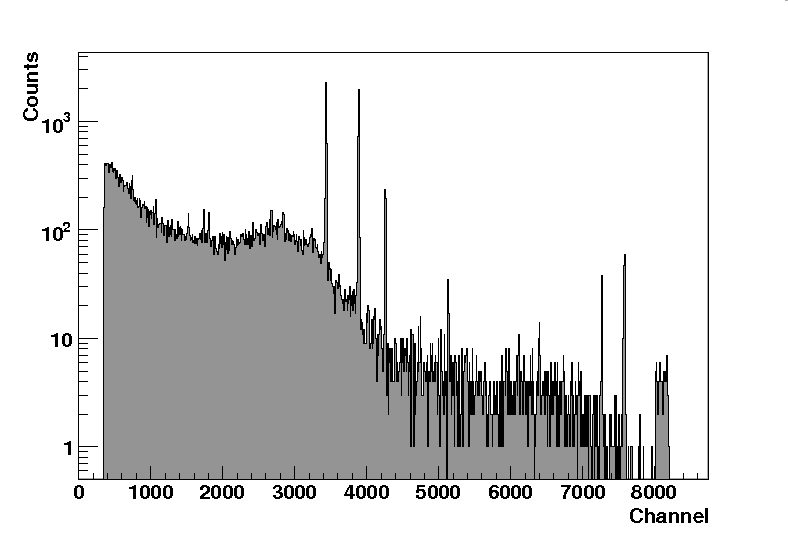}
\includegraphics[width=0.49\textwidth]{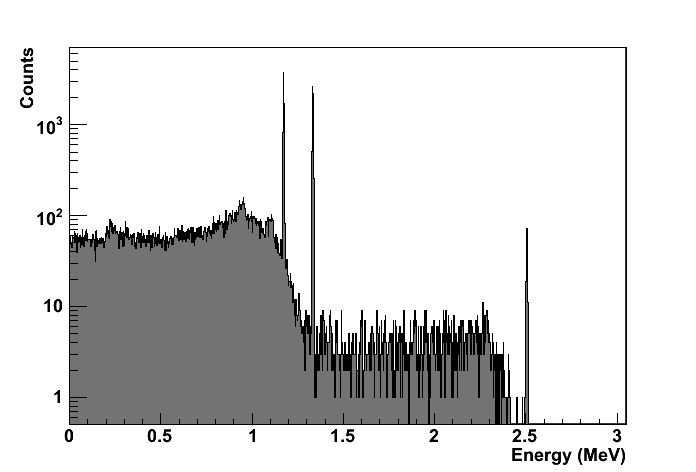}
\includegraphics[width=0.49\textwidth]{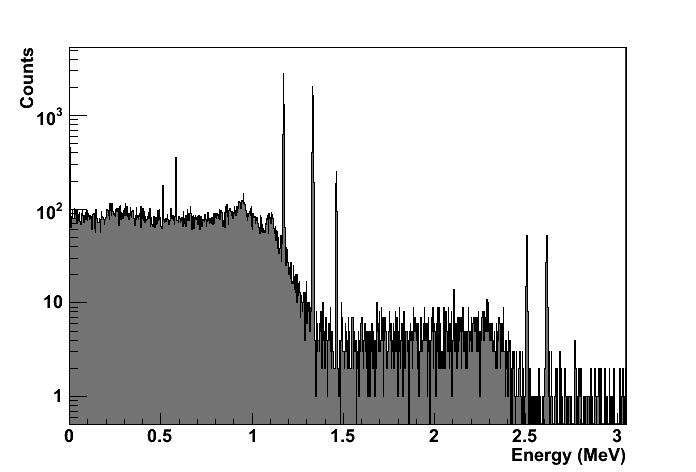}
\caption{A typical spectrum from the physical HPGe detector (top) and from the Monte Carlo simulation without background (middle) and with background (bottom). Note the two $^{60}$Co lines at 1.17~MeV and 1.33~MeV in all spectra, a $^{40}$K peak at 1.46~MeV for data with background, a $^{60}$Co sum peak at 2.51 MeV in all spectra, and a $^{208}$Tl peak at 2.61~MeV for the spectra with background. The real data shows a saturation effect past channel~8000.}
\label{fig:samplespectrum}
\end{figure}

We obtained the $\gamma$-ray energy spectra of a $^{60}$Co point source at different positions around the detector.  A flat piece of acrylic was taped to the detector cryostat can to hold the source at a set location. We moved the source along a grid with a 2~mm spacing that was marked on the acrylic sheet, as shown in Fig.~\ref{fig:detectordiagram}. The source was started 5~mm from the detector endcap, and data points were taken until the source was approximately 35~mm away from the detector. Each data point was taken for 45~seconds, which yielded an average of 80~000 counts per spectrum. We also varied the exposure time in 5-second intervals from 5 to 45~seconds for a location 5~mm from the outer casing of the detector. This data was used to quantify statistical effects on the determination of the position of the $^{60}$Co source. An MCA acquired an energy spectrum from the HPGe detector during each run. As shown in Fig.~\ref{fig:samplespectrum}, the major peaks identified on the spectra were the two $^{60}$Co peaks, at 1.173~MeV and 1.332~MeV, a background $^{40}$K peak at 1.460~MeV, a $^{60}$Co sum peak at 2.505~MeV, and a background $^{208}$Tl peak at 2.614 MeV. $^{208}\mathrm{Tl}$ is a daughter isotope of naturally-occurring $^{232}\mathrm{Th}$.

\begin{figure}
\center{
\includegraphics[width=0.49\textwidth]{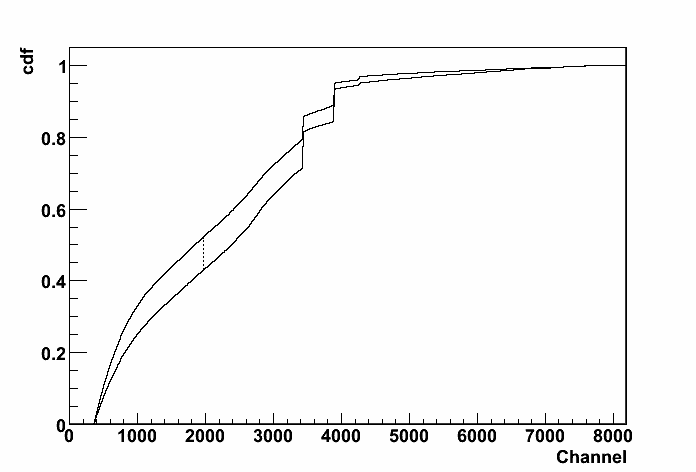}
\includegraphics[width=0.49\textwidth]{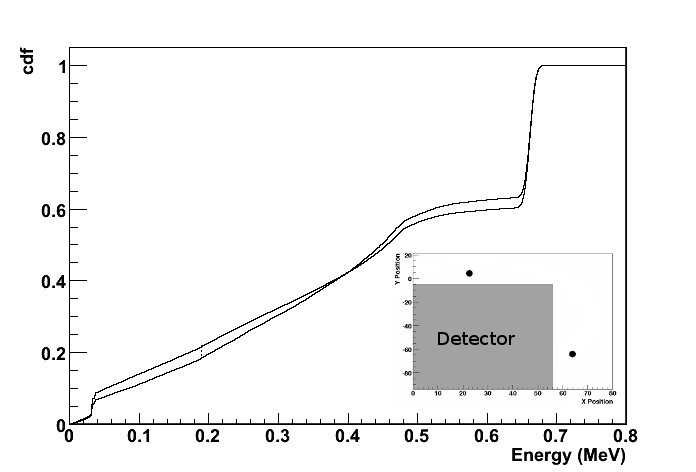}
}
\caption{Cumulative distribution functions for two physical $^{60}$Co spectra taken at different locations (top) and two simulated $^{137}$Cs spectra taken at the same locations (bottom). The $^{137}$Cs spectrum shown here was simulated without background. The dotted vertical lines show the bins with the largest difference between the two cumulative distribution functions. In both instances the most sensitivity to differences in the distribution was in the cumulant of the low energy half of the Compton continuum. The diagram in the bottom right of the right plot shows the locations of the two positions being compared in both plots. The K-S test value between these two points was equal to zero within machine precision in both cases.}\label{fig:cdf}
\end{figure}

\begin{figure}
\centering
\includegraphics[width=0.49\textwidth]{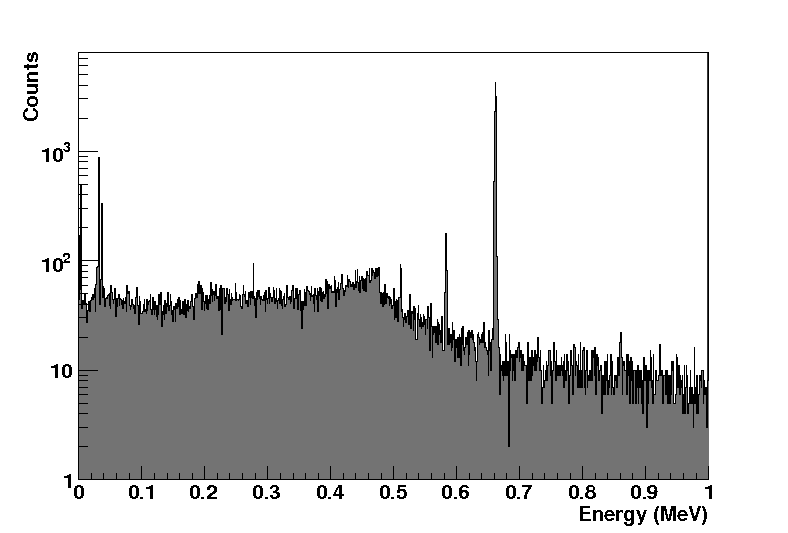}
\caption{A simulated $^{137}$Cs spectrum with backgrounds. Note the 0.662~MeV peak from the $^{137}$Cs, as well as the background $^{208}$Tl peak at 0.583~MeV. The same amount of background as Fig.~\ref{fig:samplespectrum} was used here.}\label{fig:cssamplespectrum}
\end{figure}

Analysis programs were coded in C++ and in ROOT~\cite{brun} to compare a spectrum acquired at an arbitrary location (called a \emph{test spectrum}) with the spectra taken at all locations (called the \emph{reference spectra}). Specifically, it applied a well-known goodness-of-fit test, the Kolmogorov-Smirnov (K-S)  test, between a test spectrum and all the reference spectra. The K-S test finds the maximum distance between two cumulative distribution functions, in our case the cumulants of the energy spectra. This maximum distance is inversely correlated to the likelihood that the two distributions came from the same distribution, and is converted into a probability using the Kolmogorov distribution~\cite{kanji}. This version of the K-S test is ideally used on unbinned data, but with binned data the output of the test is still a reliable measure of how similar the two distributions are if the bin-size is smaller than the energy resolution, as is the case here. We chose the K-S test as it is sensitive to subtle differences in the shape of spectra, conceptually simple, and computationally fast. We found that the logarithm of the K-S test value was a better indicator than the actual value of how correlated two spectra were. Typical cumulative distribution functions used in the study are shown in Fig.~\ref{fig:cdf}. Interestingly, these results indicate that the middle of the Compton continuum provides the K-S test with the most sensitivity to position-dependent differences in the spectral-shapes. It is less sensitive to the peak-to-Compton ratio, for example. This may also be a manifestation of the K-S test's propensity to be less sensitive to the tails of distributions. Regardless, it still indicates that there is position-sensitive information contained in the spectral shape of the low-energy half of the Compton continuum. 

A Monte Carlo simulation in GEANT4 \cite{agost} was used to simulate this experiment as well. The simulation geometries were created to match the actual experiment as closely as possible, and $^{60}$Co and $^{137}$Cs sources were simulated around the HPGe detector at the same locations the $^{60}$Co radiation source was measured around the physical detector. Typical simulation spectra are shown in Figs.~\ref{fig:samplespectrum} and ~\ref{fig:cssamplespectrum}. A Gaussian energy smearing was applied to include the effect of finite detector energy resolution. $^{40}$K and $^{208}$Tl backgrounds were also simulated. The intensity of the background in the simulations was varied to study the effects, as described later.

\section{Results}

\subsection{Peak Analysis}

\begin{figure}
\center{
\includegraphics[width=0.44\textwidth]{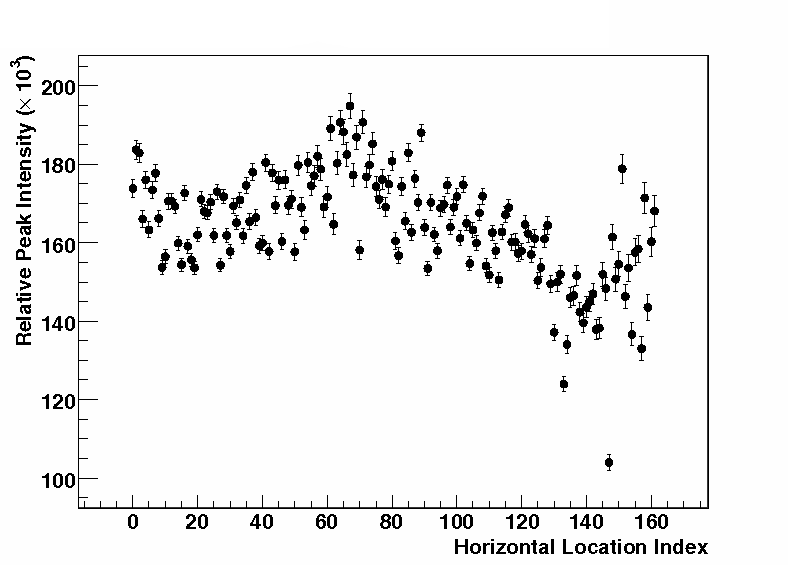}
\includegraphics[width=0.44\textwidth]{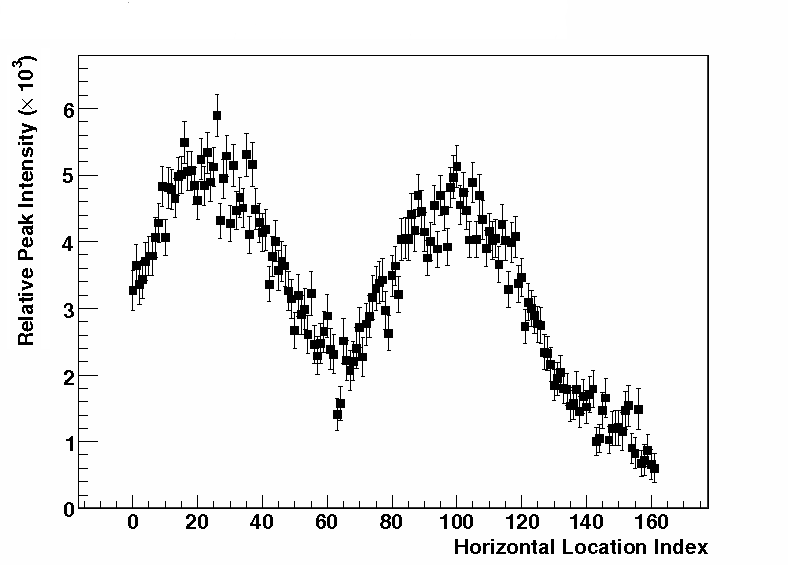}
\includegraphics[width=0.44\textwidth]{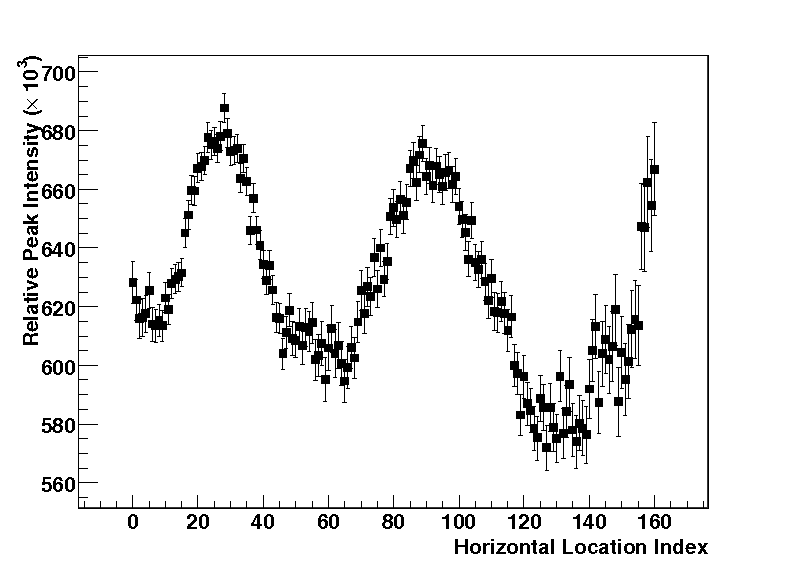}
}
\caption{Change in the $^{60}$Co 1.33~MeV relative peak height (top), the relative $^{60}$Co sum peak height (middle), and the $^{137}$Cs 662~keV relative peak height (bottom) as a function of source position. Relative peak intensity refers to the peak area divided by (the total area minus the peak area). The source was moved in incremental steps across the face of the detector and then along the side of the detector, maintaining a distance of 30~cm away from the endcap. Each step corresponds to an increment in the horizontal location index on the graphs' $x$-axes. The sharp turns in the middle of the bottom plots show where the source was turned from the front of the detector to the side. We notice that little variation exists in the $^{60}$Co 1.33~MeV relative peak height, but there is great variation in the relative sum peak heights, as well as the $^{137}$Cs 662~keV relative peak height. Errors shown are statistical.}
\label{fig:AreaChange}
\end{figure}

We performed a quick analysis to verify our analysis against earlier work concerning the peak-to-compton ratio. Fig.~\ref{fig:AreaChange} shows the change in the 1.33~MeV peak area and the sum peak area relative to the total area as a function of position around the detector for physical $^{60}$Co measurements. Obviously, there is little variation in the relative 1.33~MeV peak area, but a clear trend is seen with the sum peak. When the radiation source was placed at the center of the front of the detector and halfway down the side of the detector, the relative sum peak area was roughly five times greater than when the radiation source was placed behind the detector. This is due to the fact that the solid angle subtended by the detector is greatest at the center of the front of the detector and the center of the side, as previously described in~\cite{prata}. This difference was clearly visible without conducting statistical tests. The 1.17~MeV peak area differed with roughly the same distribution as the 1.33~MeV peak, with a difference in amplitude because the two peaks depend on the solid angle subtended by the detector in the same manner~\cite{knoll}. 

Fig.~\ref{fig:AreaChange} shows the change in the 662~keV peak area relative to the total area as a function of position around the detector for simulated $^{137}$Cs data. Unlike the 1.17~MeV and 1.33~MeV $^{60}$Co peaks, a clear trend is visible.

\subsection{Statistical Analysis with the Kolmogorov-Smirnov Test}
\begin{figure*}
\centerline{
   \mbox{\includegraphics[width=0.34\textwidth]{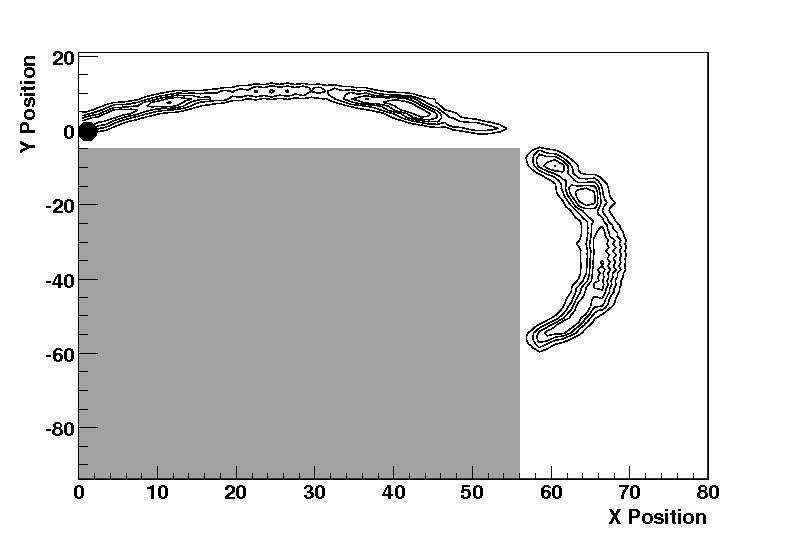}}
   \mbox{\includegraphics[width=0.34\textwidth]{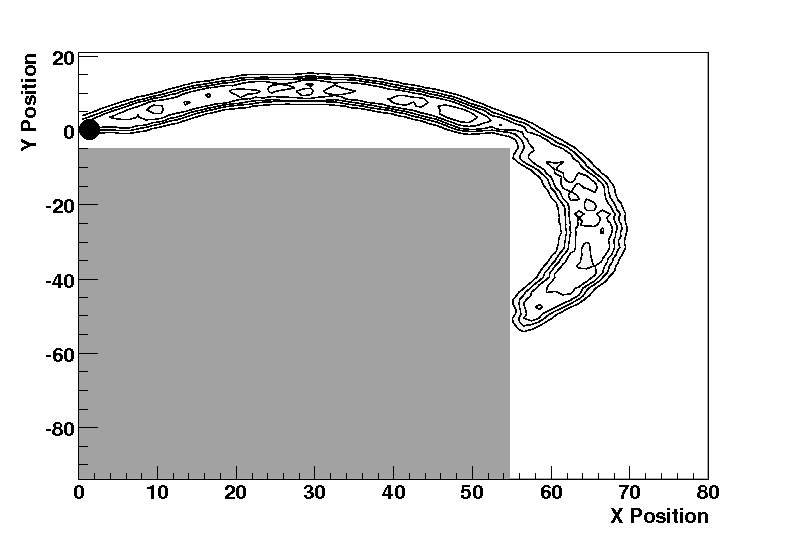}}
}
\centerline{
   \mbox{\includegraphics[width=0.34\textwidth]{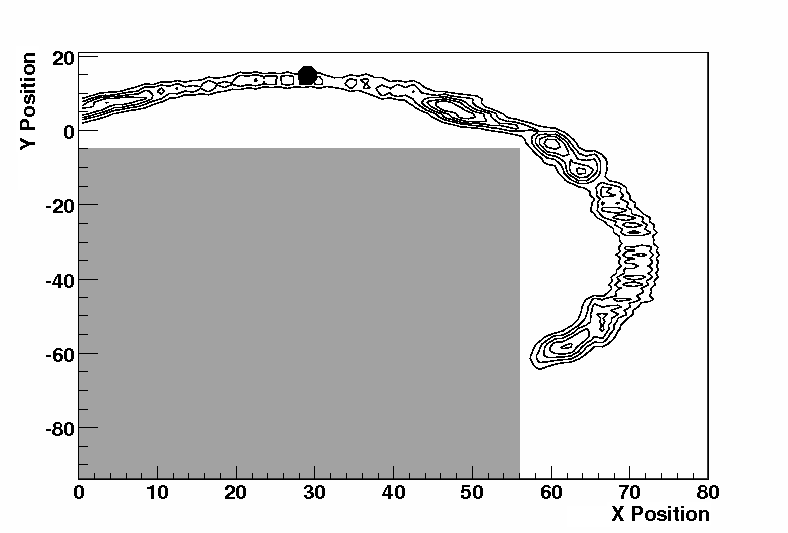}}
   \mbox{\includegraphics[width=0.34\textwidth]{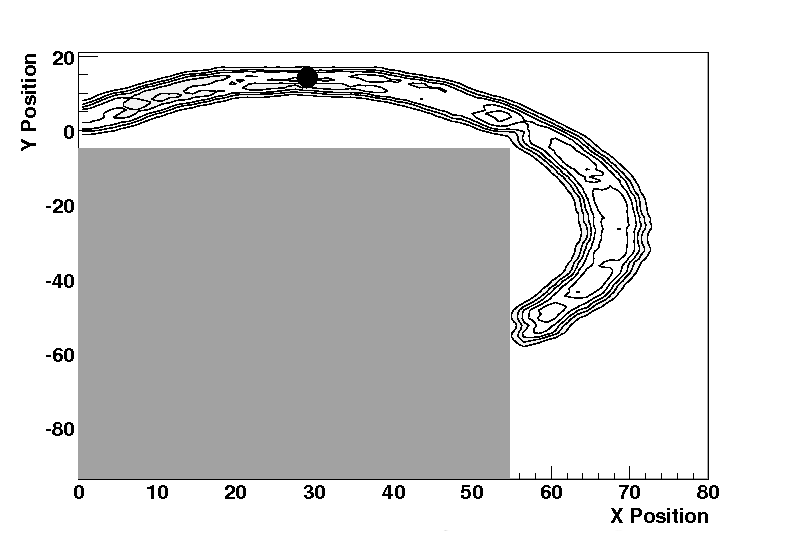}}
}
\centerline{
   \mbox{\includegraphics[width=0.34\textwidth]{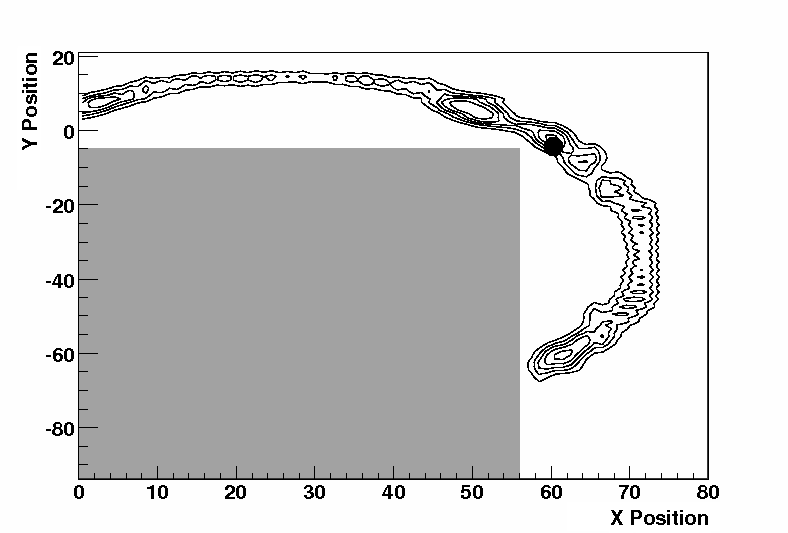}}
   \mbox{\includegraphics[width=0.34\textwidth]{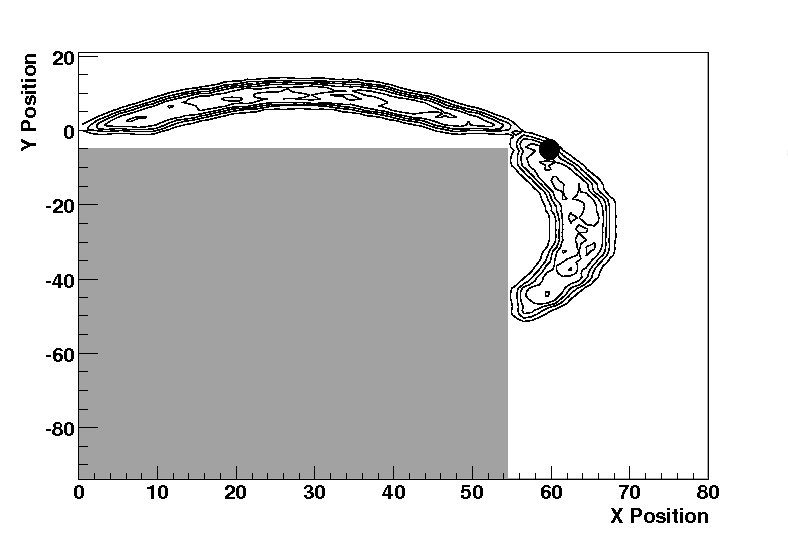}}
}
\centerline{
   \mbox{\includegraphics[width=0.34\textwidth]{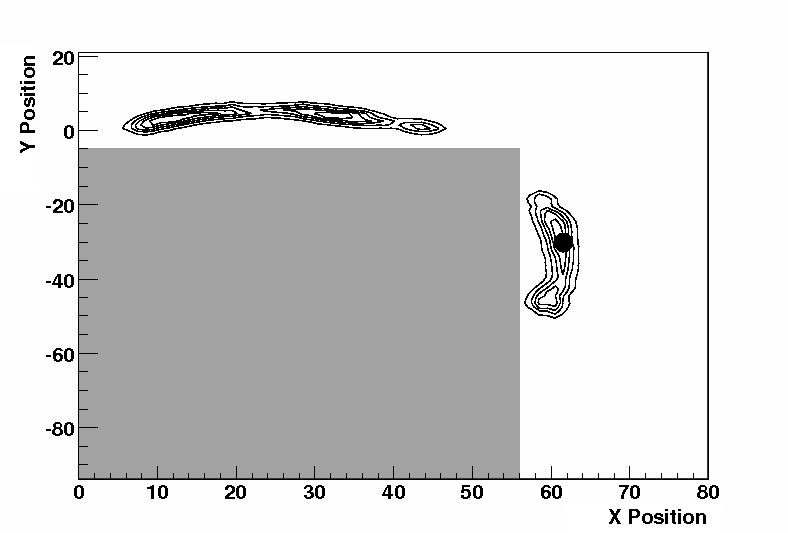}}
   \mbox{\includegraphics[width=0.34\textwidth]{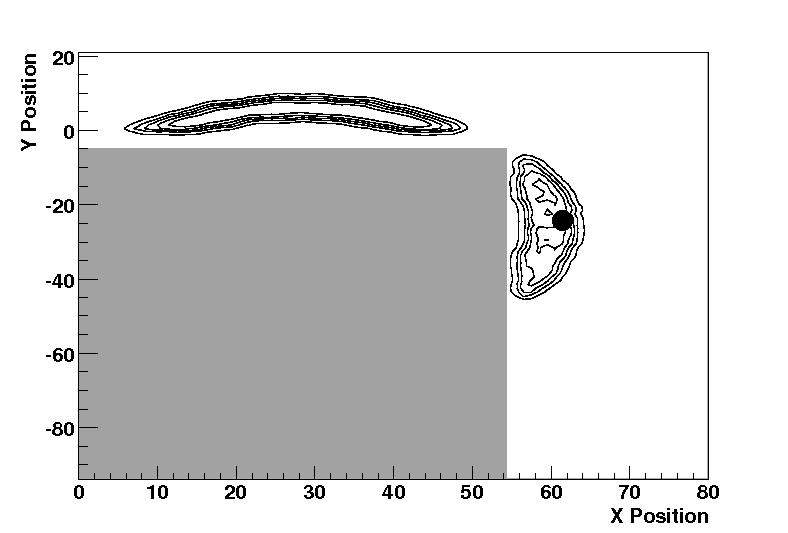}}
}
\centerline{
   \mbox{\includegraphics[width=0.34\textwidth]{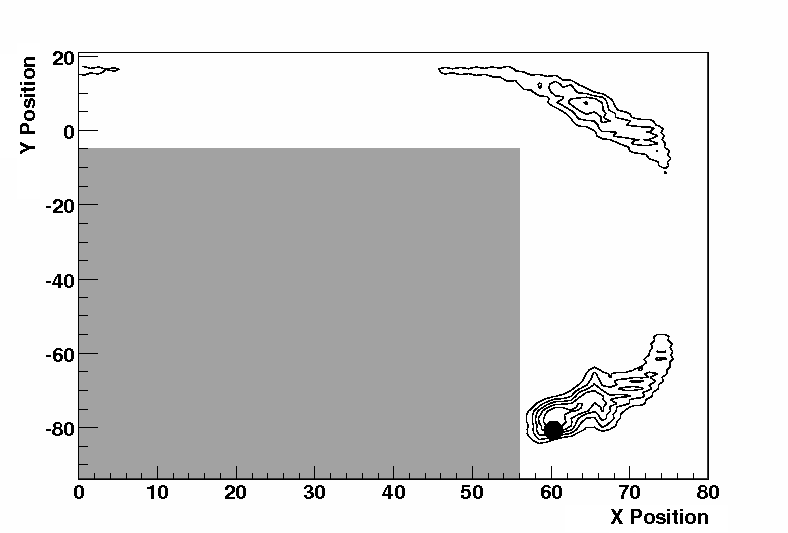}}
   \mbox{\includegraphics[width=0.34\textwidth]{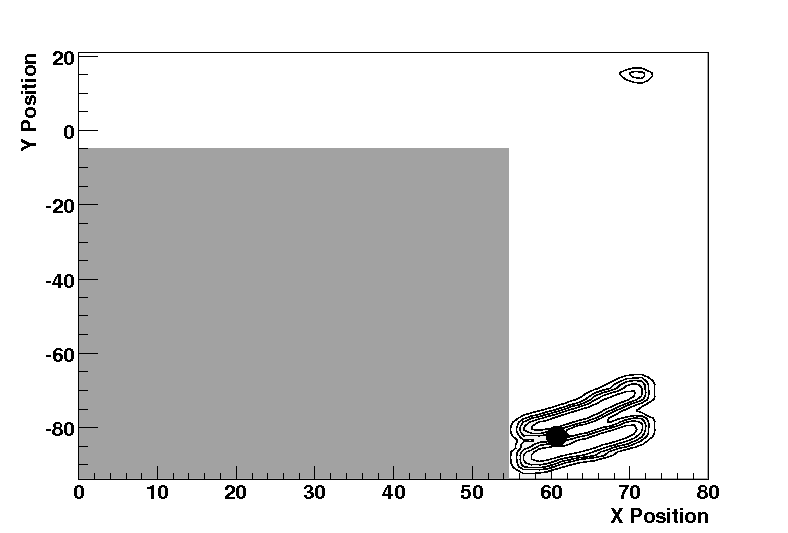}}
}
\caption{Contour maps of K-S test values between the test spectrum and reference spectra for the actual measurements (left column) and the Monte Carlo simulation (right column) for a $^{60}$Co source. The test source position is shown with a black dot. The white areas outside of contours represent areas where the K-S test value between that point and the black dot is less than~$10^{-55}.$ The detector is located at the gray area in the lower left of the diagram. Each graph has been smoothed, and each contour in this plot represents a difference in the logarithm of the K-S test value of 9.}\label{fig:ReconstructionArch}
\end{figure*}

\begin{figure}
\center{
\includegraphics[width=0.46\textwidth]{Heatmap14.png}
\includegraphics[width=0.46\textwidth]{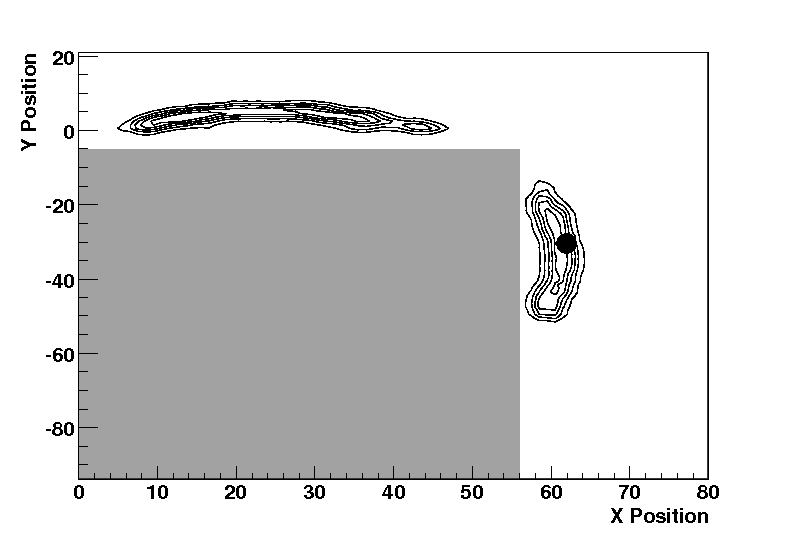}
}
\caption{Contour maps of the measured values of the K-S test value between the spectrum at each position and the test source spectrum for the actual measurements using the entire $^{60}$Co spectrum (left) and bins up to, but excluding, the 1.17~MeV $^{60}$Co peak (right). Although significant changes were seen in the sum peak height, the contour plots are basically indistinguishable from one another, indicating that the Compton continuum has significant information about the source position as well.}\label{fig:NoSumPeak}
\end{figure}

\begin{figure}
\centering
\includegraphics[width=0.45\textwidth]{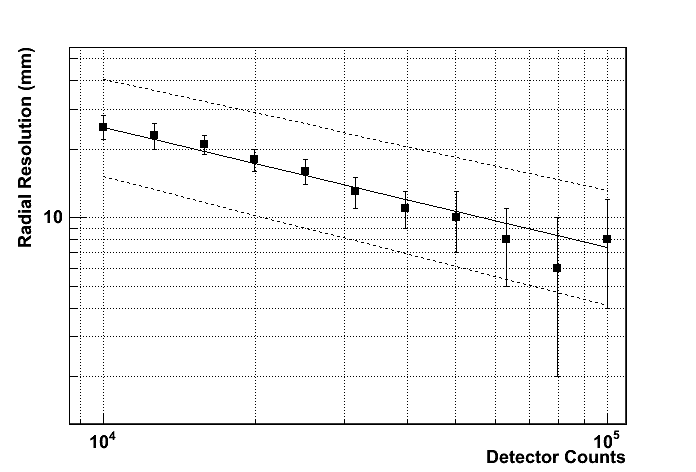}
\includegraphics[width=0.45\textwidth]{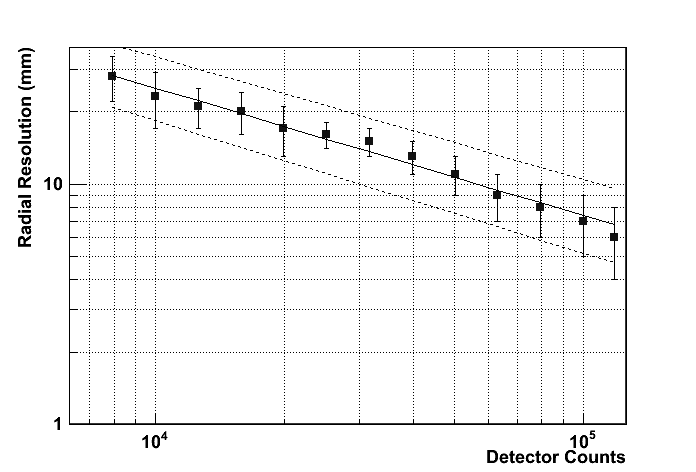}
\includegraphics[width=0.45\textwidth]{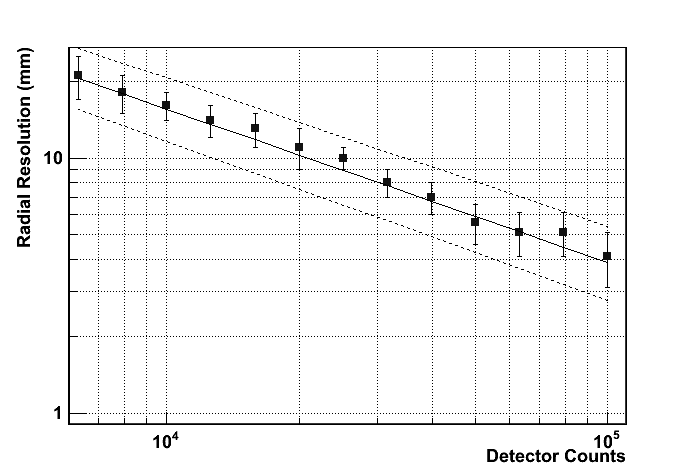}
\caption{Spatial resolution as a function of total detector counts for detector measurement $^{60}$Co (top), Monte Carlo $^{60}$Co (middle) and $^{137}$Cs (bottom) data without background. We defined radial resolution as the width of the region where the logarithm of the K-S test value was greater than -5. The error was defined as the region where the logarithm of the K-S test value was greater than -10. The best fit line (solid line) and the upper and lower bound fits (dotted lines) are shown. For $^{60}$Co, in both detector measurements and Monte Carlo simulations the resolution is proportional to the $-0.53\pm 0.02$ power of detector counts, and the constant of proportionality is $3.3\pm0.5$~m for detector measurements and $3.3\pm0.4$~m for Monte Carlo. For $^{137}$Cs, the resolution is proportional to the $-0.60\pm 0.02$ power of detector counts, and the constant of proportionality is $3.9\pm0.4$~m}\label{fig:Resolution}
\end{figure}

\begin{figure}
\centering
\includegraphics[width=0.49\textwidth]{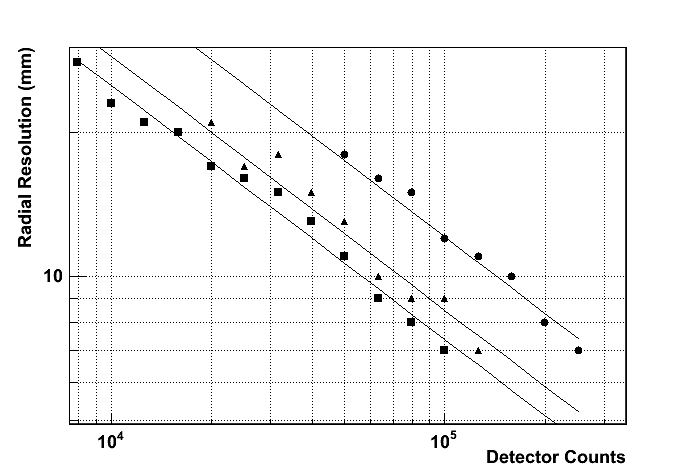}
\includegraphics[width=0.49\textwidth]{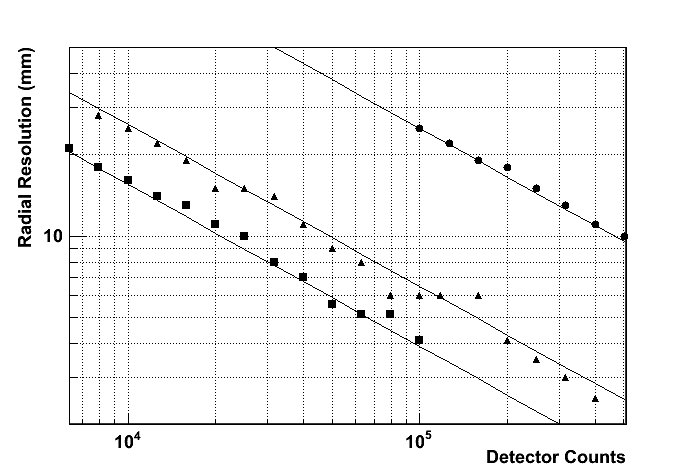}
\caption{Best fit lines with data points for resolution of spatial reconstruction as a function of total detector counts for differing amounts of background for Monte Carlo $^{60}$Co (top) and $^{137}$Cs (bottom) data. In the $^{60}$Co plot, the bottom line (with square data points) has no background (as in Fig.~\ref{fig:Resolution}), the middle line (with triangular data points) has a ratio of $^{40}$K peak area to 1.17 MeV $^{60}$Co peak area of 0.25, and the top (with circular data points) has ten times more background counts the middle one. In the $^{137}$Cs plot these proportions are the same, using the 0.583 MeV $^{208}$Tl as the background peak instead of $^{40}$K, and the 0.662 MeV $^{137}$Cs peak instead of the 1.17 MeV $^{60}$Co peak. For the highest background rate, the fit was extrapolated from higher detector counts, since our simulation was not sensitive to resolutions greater than 30~mm. With an increase in background, the proportionality to detector counts in the power law does not change, but the constant of proportionality increases.}\label{fig:bgres}
\end{figure}

\begin{figure}
\centering
\includegraphics[width=0.48\textwidth]{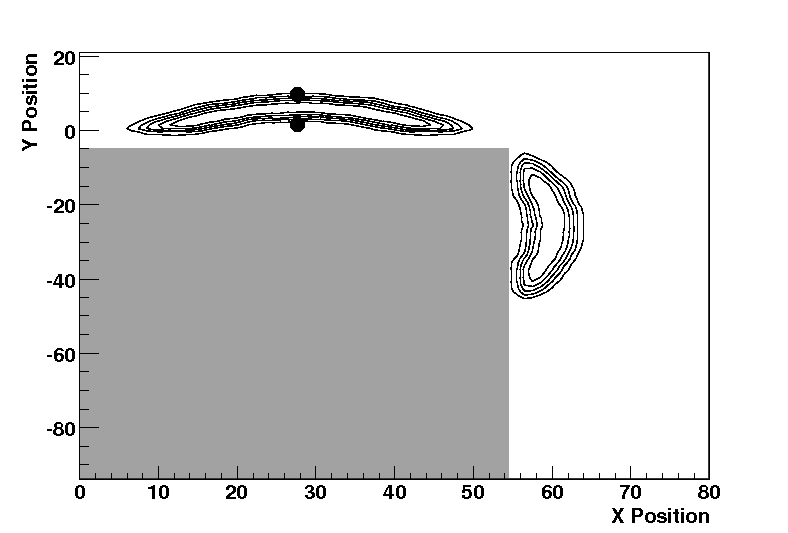}
\caption{Contour map of the measured values of the K-S test value between the spectrum at each position and the superposition of two test source spectra for simulated Monte Carlo data. The location of the two sources of equal strength are shown. Note that adding an extra source causes the contours to expand.}\label{fig:TwoSrc}
\end{figure}

Fig.~\ref{fig:ReconstructionArch} shows the contour plots of the K-S test values for four different test spectra compared to the reference spectra for physical and simulated $^{60}$Co data. The resulting regions with high K-S values appear like rings on the plots, corresponding to approximately spherical shells in three dimensions. As the figure shows, the areas of high K-S test values do not necessarily point out the angular position of the source. Fig.~\ref{fig:ReconstructionArch} also indicates good agreement between data and Monte Carlo and gives us confidence in the simulated $^{137}$Cs results. We note there is a interesting 'double-arch' structure in the bottom-most right plot of Fig.~\ref{fig:ReconstructionArch}. This is likely due to picking a test point that lies between two reference spectra points. As shown in Fig.~\ref{fig:NoSumPeak}, a similar plot was made comparing the K-S test values for histograms filled up to (but not including) the 1.17 MeV peak in the $^{60}$Co spectrum. These plots were barely distinguishable from those when all peaks and overflow bins were included, showing that although the difference in the sum peak area relative to position was quite noticeable, the Compton continuum also has significant influence on the differences between the spectra. Similar results were seen with the simulated $^{137}$Cs data. This is consistent with our earlier observations of the dominant effect of the low-energy half of the Compton continuum on the K-S value. 

Using test spectra from 20 randomly-selected points, we found that the radial position of the radiation source could be found to within 5 mm 95\% of the time, yet the angular position could be found to this resolution only 20\% of the time. We found that higher statistics resulted in lower K-S test values between the test spectrum and reference spectra for reference spectra that were further apart radially. This is shown in Fig.~\ref{fig:Resolution}. For $^{60}$Co, we found the resolution to be well-described as a power-law that is proportional to a $-0.53\pm 0.02$ power of detector counts, indicating the statistical dependence of the resolution. For $^{137}$Cs, the resolution is better described by a power-law that is proportional to a $-0.60 \pm 0.02$ power of detector counts. As shown in Fig.~\ref{fig:bgres}, with an increase in background, the resolution is still described by a power law with the same power of detector counts, but the absolute resolution increases as the number of background events increases, as expected. Nor surprisingly, this indicates that one can still improve the sensitivity of the radial resolution in the case of large backgrounds by collecting more statistics. 


When the test spectrum is a spectrum that is the superposition of spectra from two distinct sources at different positions, the rings of Fig.~\ref{fig:ReconstructionArch} widen, with the amount it widens determined by how far apart the two sources are. These rings are in between where the rings for each individual source would be, with the exact location determined by the difference in intensity of the two sources. This is shown in Fig.~\ref{fig:TwoSrc}.

\section{Conclusions}

We were able to accurately reconstruct a bright, nearby radiation point source's radial position using an array of reference spectra and the K-S test. We were also able to accurately reproduce these results using Monte Carlo simulations, indicating that this method can be used by comparing detector spectrum to sets of simulated spectra. The K-S test appears to be most sensitive to spectral shape effect in the Compton-continuum. However, using our method we were not able to gain good angular position information about a radiation source. Future work will determine the limits of this techniques using unbinned data and low statistics. Low background HPGe detector arrays, such as {\sc Majorana}, can potentially use this technique along with relative rates in detectors to determine the location of unwanted hot-spots in the detector array that were inadvertently introduced during construction. This technique is also applicable to other gamma-ray spectrometers, such as NaI and cryogenic bolometer detectors. We also want to point out one note of caution with using the K-S test in the presence of sharp peaks. The K-S test would indicate incompatible distributions when comparing spectra with peaks that are slightly offset from each other, ie. when there exist slight binning misalignments, calibration offsets, or simulation-vs.-data errors. Care must be exercised the ensure the proper calibration and alignment of the spectra's histogram bins.

\section{Acknowledgements}
We would like to thank the LENA group and Prof. A.E.~Champagne for the use of the LENA HPGe detector. This work is supported under in part by DOE NP Grant \# DE-FG02-97ER41041 and the State of North Carolina.

\end{document}